\documentclass[aps,preprint,groupedaddress]{revtex4-2}

\usepackage{multirow}
\usepackage{array}
\usepackage{amsmath}
\usepackage{amsfonts}
\usepackage{graphicx}
\graphicspath{{./}}
\usepackage{url}
\usepackage{hyperref}

\usepackage[table]{xcolor}
\usepackage{tabularx}
\definecolor{lightgray}{gray}{0.95}
\usepackage{booktabs}
\usepackage{enumerate}
\usepackage{longtable}
\usepackage[english]{babel}
\usepackage{dcolumn}
\usepackage{bm}

\begin{document}

\title{Constraints From Simulation Improve Experiential Outcomes in Laboratory Environment}

\author{Jeff Bale}
\author{James Day}
\email[Corresponding author: ]{jday@phas.ubc.ca}
\author{J. Ives}
\author{D. A. Bonn}
\affiliation{Department of Physics and Astronomy, University of British Columbia, Vancouver, British Columbia, Canada V6T 1Z1}

\date{\today}

\begin{abstract}
In a first-year physics inquiry lab, pairs of students were randomly assigned to study pendulum motion using either a physical apparatus or a computer simulation. The experiment required detecting a ${\sim}1\%$ difference in period between pendulums released at $10^\circ$ and $20^\circ$. This is the subtle failure of the small angle approximation and a goal that demands iteratively refined, high-precision measurements. Students using the simulation achieved significantly more reproducible timing measurements across all rounds of data collection and, by the third round, had also adopted more effective data collection strategies overall. As a result, 78\% of simulation groups met the precision threshold required to identify the model failure, compared with 52\% of physical apparatus groups. We attribute these outcomes primarily to a specific simulation constraint: students were required to use the simulation's built-in timer, which forced them to decouple pendulum release from the start of timing. This prevented students from pursuing a reaction-time-limited synchronization strategy that often traps users of physical apparatus in a low-precision measurement dead end. A post-lab survey further shows that students using simulations were more confident in their results than those who instead used a physical pendulum, as well as preferred greater use of simulations in future labs. These findings suggest that carefully designed simulation constraints can guide students toward productive experimental strategies while preserving their investigative autonomy.
\end{abstract}

\maketitle

\section{\label{Sec:Intro} Introduction}

Numerous studies have explored how simulations fare against their physical counterparts in laboratory courses ~\cite{Finkelstein2005,Klahr2007,Zacharia2008,Zacharia2011,Jariwala2019,Kapici2019}. A comprehensive review indicates that a significant majority of these studies (n~=~50, 89\%) show that student learning outcomes in non-traditional (simulated and remote) labs are comparable to, if not better than, those in traditional (hands-on and in-person) settings across various learning outcome categories ~\cite{Brinson2015}. While the majority of these studies have concentrated on disciplinary content knowledge, less attention has been paid to analytical and inquiry skills. These latter categories, which encompass the capacity to observe, hypothesize, design experiments, and understand the nature of science, as well as to critique, interpret, and integrate experimental data, are central to our investigation. Over recent years, our focus has been on developing first-year physics labs whose central goal is to enhance students' data-driven reasoning abilities, primarily targeting these inquiry and analytical skills ~\cite{Holmes2014}. The lab course has no explicit goals in relation to specific physics content or concepts. Instead, it concentrates on analytical and inquiry skills, using a framework based on comparative analysis, starting with simple measurements and gradually advancing to complex comparisons between data and theoretical models, using weighted least squares. The course structure grants students significant autonomy in making measurement decisions throughout the laboratory activities, encouraging iterative improvements and bolstering confidence in their experimental results.

In this study, we examine the comparative effectiveness of conducting this type of inquiry lab with computer simulations versus using traditional physical apparatus in the same in-person environment. Given the adaptability of simulations for online learning and their potential to enhance educational accessibility, our study aims to compare the impact of these two approaches on student engagement and learning outcomes. We investigate their inquiry skills by evaluating their experimental choices and the quality and precision of their data. Their analytical skills are examined by looking at their iterative improvements and how they draw conclusions from their overall dataset.

We find evidence that affordances and constraints in the simulation enable students to gather higher-precision data with better measurement strategies. The carefully designed simulation gives them a greater chance of detecting a subtle model failure, the main goal of this experiment. Furthermore, we provide an analysis of student feedback suggesting a somewhat improved experience for students who use simulations.

\section{\label{Sec:Context_and_Data} Context and Data Collection}

\subsection{\label{SubSec:Course} Course context}
Data were collected from a stand-alone physics laboratory course offered by the University of British Columbia in Vancouver, Canada. Participants in this study consisted primarily of general science majors who were also taking a calculus-based first-year physics course, though a minority planned to pursue a degree in physics. The students' data and survey information were collected over one week during an academic term running from January to April of 2018, during which students attended a weekly three-hour lab session. Two teaching assistants and one instructor facilitated each lab section. While the instructors and teaching assistants varied throughout the week, they all followed precise instructions to ensure consistency across the sections.

The sequence of experiments in this course is supported by the introduction of the statistical tools needed to work with the data gathered. The development of students’ authentic scientific habits was supported by an experimental cycle that led them through an iterative process~\cite{Collins1988} in which they reflected on the results of a comparison of their data, devised a means to improve their experiment, and then executed that new plan. The use of this iterative process is strongly scaffolded at the beginning of the course through the lab manual and grading rubrics, and then faded over time. Previous work has shown that this results in changes in student behaviour and improvements in the quality of students' reasoning~\cite{Holmes2015teaching}.

\subsection{\label{SubSec:Goals} Student goals and tools}

The lab activity that is the subject of this study is pivotal in the lab sequence, occurring in the third week of the course and lying at the heavily scaffolded end of the spectrum. Like all of the lab activities that we have developed, it is not simply confirmatory; it is not seeking a result that is well-known or has recently been taught to the students. Instead the pedagogical strength of the lab activity relies on the relatively subtle failure of a simplified model - the treatment of a pendulum as a simple harmonic oscillator by making a small angle approximation. This model (and its failure) is not pointed out to the students, nor do we ever provide them with the equation for the period of a pendulum. The task is posed as an empirical comparison where we understand they will have expectations about how the results will turn out. So, the challenge students undertake is to measure the period of a pendulum released from an angle of $10^{\circ}$ and compare it to the period of the same pendulum released from $20^{\circ}$. In this range, the failure of the small-angle approximation is at the 1$\%$ level, so it requires quite high-precision data to discern the model failure, but is readily achievable in the time they have available.

What makes the experiment so important in our lab sequence is that each student makes a challenging journey through seemingly contradictory outcomes. They typically begin by observing no difference in the periods, and then, ideally, they gradually uncover a disagreement. This creates a period of confusion, almost a crisis, that requires them to trust their highest-quality data. To enable the students to complete this cycle, a very careful tuning of the lab design is required. We always aim to provide free agency in these activities and enough time to succeed. However, time constraints often still conflict with our ideal of every student reaching a satisfying endpoint. Our results will show that groups using either type of apparatus will typically progress through this arc of initially observing no difference in the periods due to low-quality data, to observing a difference by their third round of data collection after iterative improvements in their experimental approaches. In the next section, we show how the affordances and constraints of a simulation can help them progress more quickly toward sufficiently high-quality data to support this conclusion.

At this stage of the course, we introduce the students to a simple tool for comparing two measured quantities with uncertainty. In order to avoid some misconceptions around measurement uncertainty~\cite{Buffler2001} this course moves away from the dichotomous scale of agree or disagree, and towards a continuous scale. Namely, how much disagreement exists between two measurements? They already know how to use mean and standard uncertainty of the mean to calculate values from a set of repeated measurements, so can produce results for two different data sets, $A\pm \delta A$ and $B\pm \delta B$. The general comparison tool that we developed for our first year labs \cite{Holmes2015} is simply the difference between the mean values $A$ and $B$, relative to the combined uncertainty in the two values. This is referred to as a t$^\prime$-score, owing to its similarity to Student's t-test~\cite{Student1908},

\begin{equation}\label{Eq:t_score}
  t^\prime \mbox{-score} = \frac{A - B}{\sqrt{(\delta A)^2 + (\delta B)^2}}.
\end{equation}

\begin{figure}[htpb]
  \centering
  \includegraphics[width=0.5\textwidth]{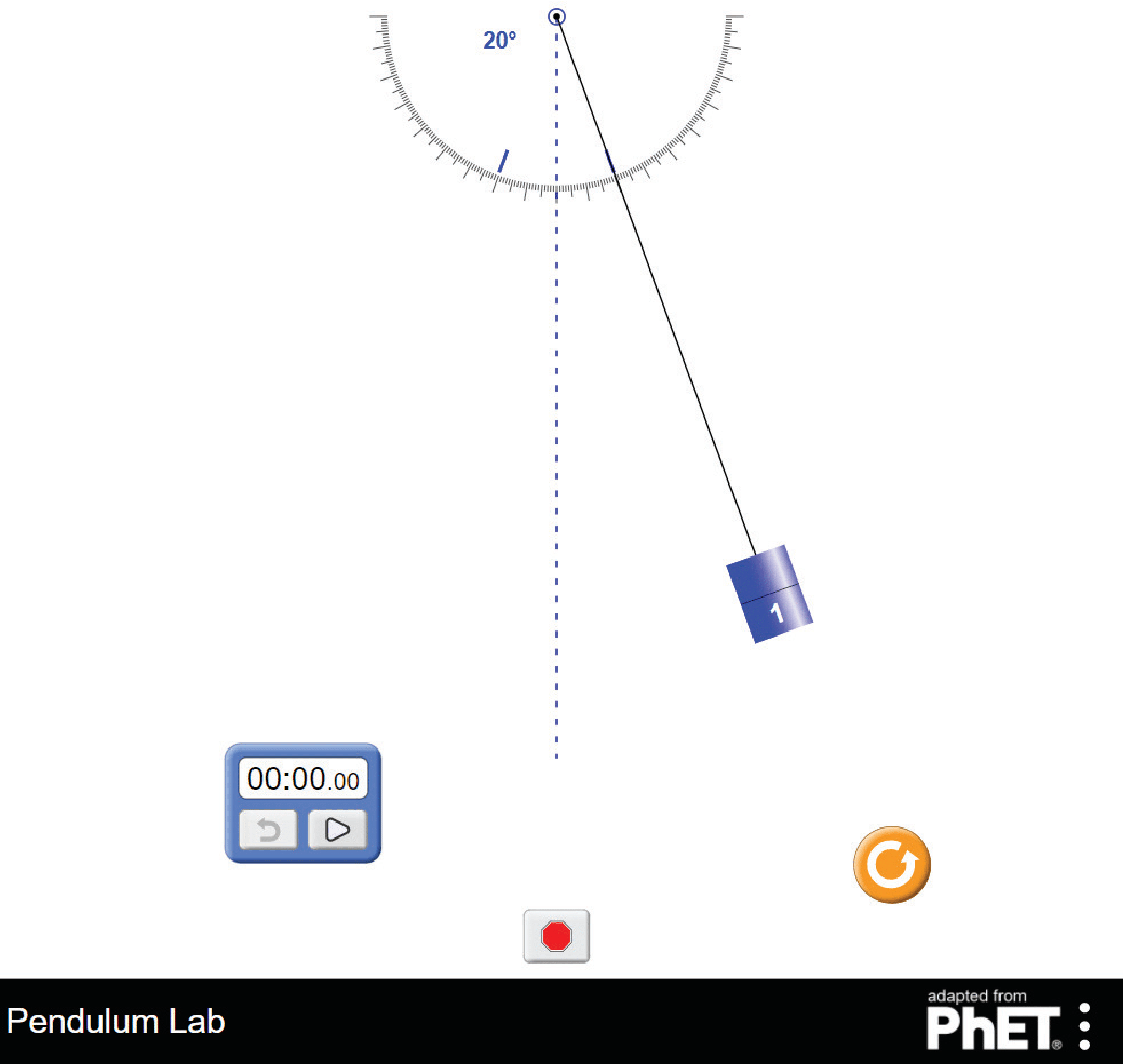}
  \caption{The student view of the modified PhET used for this study. The simulation is extremely simple and includes nothing more than a pendulum with a protractor, a stopwatch, and stop and reset buttons.}
  \label{fig:pendulum_phet}
\end{figure}

In the experiment we examine here, the t$^\prime$-score is used by the students to compare the periods measured at starting angles of $10^{\circ}$ and $20^{\circ}$. Without using a sophisticated t-test, the t$^\prime$-score offers students a continuous scale to assess disagreement between two measurements. The t$^\prime$-score interpretation that is introduced is that a score near 1 suggests indiscernible difference. Scores greater than 3 indicate clear differences, while scores between 1 and 3, a \textquotedblleft grey zone" of \emph{tension}, suggesting potential but unconfirmed difference. In all cases, students are encouraged to refine their measurements to reduce uncertainties and clarify their results. The comparison tool is connected directly to scaffolding of their lab activity since, regardless of the size of the t$^\prime$-score, students are prompted to improve their measurements in the hopes of further clarifying or supporting their result. Our desired outcome is for students to engage in a process of iterative refinement, such that even if they do not detect the model's limitations in their initial attempts, subsequent improvements in their experimental approach allow them to discern and understand the failure of the simplified model, thereby gaining confidence in their data and drawing meaningful conclusions.
After the first two weeks of conducting lab activities with physical apparatuses (not involving the pendulum), the third experiment employed either physical pendulums or a pendulum simulation crafted by the PhET team~\cite{PhetAdaptedPendulum}, see Fig.~\ref{fig:pendulum_phet}. Equipment was distributed evenly throughout the room such that a typical table would consist either of two physical apparatus groups and one simulation group or vice-versa. The task they were given was to use the t$^\prime$-score to compare the period of a pendulum released at $10^{\circ}~(T_{10})$ and at $20^{\circ}~(T_{20})$. They can measure the period $T$ by measuring the total time $t$ that it takes for multiple periods ($M_{cycles}$) and can perform this measurement for multiple trials ($N_{trials}$) in order to obtain a mean and standard deviation. The quality of their result is directly related to their choice of $M_{cycles}$ and $N_{trials}$, as well as more subtle choices that influence their uncertainty in an individual measurement of the time $t$. This timing uncertainty is estimated using the standard deviation, $\sigma(t)$, of their multiple measurements of time. Our measure of the quality of their experimental choices is $M_{cycles}\sqrt{N_{trials}}$ since these factors come together in the equation they had for uncertainty in the mean of a measured pendulum period

\begin{equation}\label{Eq:Unc_T_Ave}
  \delta(T_{Ave})=\frac{\sigma(t)}{M_{cycles}\sqrt{N_{trials}}}.
\end{equation}

\noindent The actual experimental procedure followed was left to each group to decide, and the scaffolding led them to reflect on their results and improve them through multiple rounds of data-taking, $L_{rounds}$.

\subsection{\label{SubSec:ConAffs} Constraints and Affordances}

Within our pendulum activity, both the physical equipment and the simulation have a variety of affordances and constraints~\cite{Norman1999, Podolefsky2013}. The most important affordance, designed into both versions, is very low damping, so that the students can in principle measure very large numbers of cycles to attain high precision. Both situations also have a protractor for measuring the release angle. The simulation has the advantage of both a digital readout of angle when initiating the pendulum swinging, and then the digital one disappears, but leaves an analog scale remaining in view.

While the affordances of the simulation were designed to replicate the physical pendulum as closely as possible, the interesting part is the extra constraints. The pendulum in the simulation is constrained to two-dimensions, so does not have the precession that develops in the physical pendulum and which leads to collisions and spoiled data, and which somewhat constrains how many cycles they can measure. A subtle but significant constraint is that while those using the physical pendulum are able to both release the pendulum and start a timer simultaneously, the students with the simulation are explicitly instructed to use the built-in timer for the data collection to be accurate. (The internal timer is tied into the CPU and is internally consistent, but not externally consistent with an independent timer). This internal timer needs to be started and stopped by a mouse click, and the mouse is also needed to \textquotedblleft grab" and \textquotedblleft release" the pendulum. This forces the students using the simulation to first start it swinging, and then subsequently pick an instance to start the timer and begin counting cycles.

\subsection{\label{SubSec:Data} Data collection}

Our first source of data for the study was all of the numbers used by students using either the physical apparatus (\textit{Phys}) or the simulation (\textit{Sim}). This included the round being conducted, the number of pendulum cycles ($M_{cycles}$) per trial, and the total number of trials ($N_{trials}$) for each round of data collection. It also included all of their measurements of the time for $M_{cycles}$, at starting angles of $10^\circ$ and $20^\circ$ ($t_{10}$ and $t_{20}$), and their reported t$^\prime$-scores. These data provide information on their measurement decisions and the quality of their measurements as they progress.

The second data source is an anonymous post-lab survey (Appendix~\ref{Appendix:Supplementary Materials}). The survey was composed of both open-response and Likert-style questions~\cite{Boone2012}. This data source has a stronger focus on both the student's interpretation of their data, as well as their perspectives on the experiment as a whole.

\section{\label{Sec:Results} Results And Analysis}

\subsection{\label{SubSec:t_compare} Students' t$^\prime$-score comparisons}

\begin{figure}[htpb]
  \centering
  \includegraphics[width=0.85\textwidth]{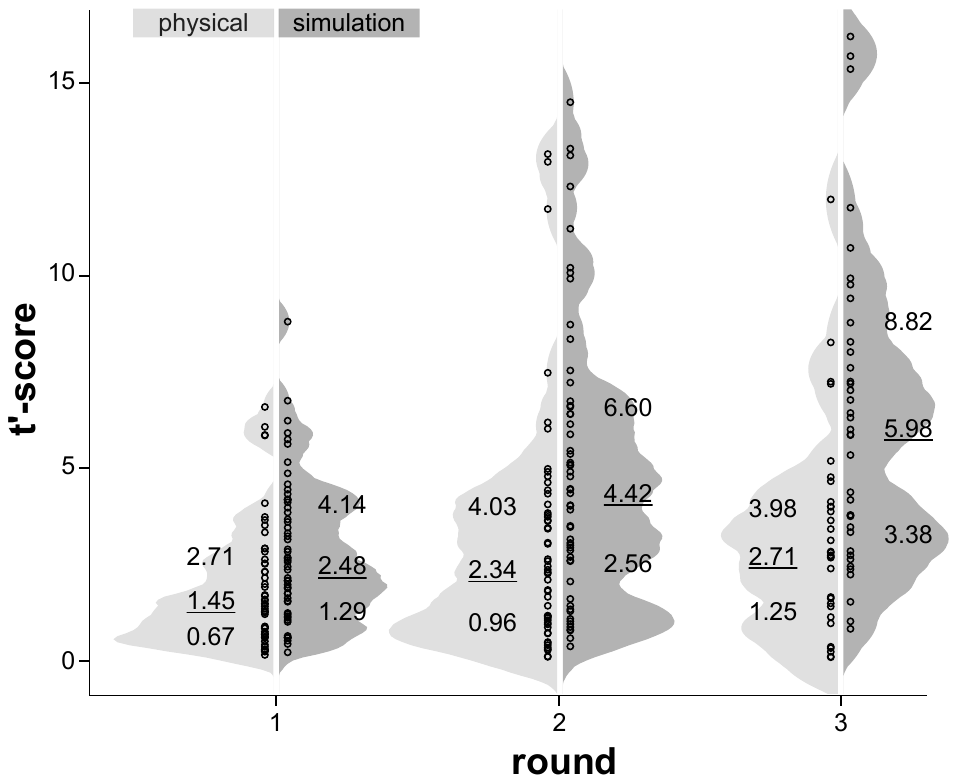}
  \caption{Evolution of t$^\prime$-scores as a function of round for both groups, illustrated using bean plots. These plots combine features of violin plots, showing the distribution of data as smooth density curves, with scatter plots that mark individual data points. This method highlights both the common values and the dispersion within the dataset, effectively visualizing the progression of data across rounds. Interpolated medians (underlined) and quartile values are superimposed on the plots to highlight the difference between the two conditions. Note: not all groups completed a third round.}
  \label{fig:tscore_evolution}
\end{figure}

A central goal of the lab activity is for students to progressively refine their experimental techniques, thereby enabling them to detect discrepancies between pendulum periods measured at $10^\circ$ and $20^\circ$ release angles. We present these outcomes through students' reported t$^\prime$-scores, displayed in Figure~\ref{fig:tscore_evolution}, which shows that students using simulations generally achieve higher t$^\prime$-scores, the key comparison needed to recognize a difference between the measured periods.

Analysis of the interpolated medians reveals a clear trend: as the rounds progress, the t$^\prime$-scores of the students using the simulation not only start higher but also increase more steeply compared to those using the physical apparatus. This places the students using the simulation in a much better position to recognize the discrepancy between the measured periods.

This difference in t$^\prime$-scores, increasing across measurement rounds, warrants a detailed examination of the factors contributing to it. The analysis below of their experimental data and design choices will highlight the mechanisms through which simulation-based learning impacts student performance and perception, and may offer insights into how different lab environments influence learning and experimentation strategies.

\subsection{\label{SubSec:Lab_data} Students' lab data}

Given the higher t'-scores reported by students using the simulation, we used a MANOVA (Multivariate Analysis Of Variance) to test if the two conditions (physical vs simulation) differ from each other, by round of data collection, in quantities such as their data-taking decisions or quality of their measurements of time. Further details on outlier handling, assumption testing, and post-hoc procedures are provided in Appendix~\ref{Appendix:Analyses_Background} to support transparency and reproducibility.

Specifically, we examined the number of pendulum cycles ($M_{cycles}$) and trials ($N_{trials}$) they chose to conduct, alongside the standard deviation of their time measurements $\sigma(t)$. The latter reflects outcomes influenced by measurement technique and apparatus constraints rather than experimental design choices or data-taking strategy. The composite quality metric $M_{cycles}\sqrt{N_{trials}}$, which affects the uncertainty in the average period, offers a combined measure of the quality of their data-taking strategy. Higher values of this metric signify a better data collection strategy as shown in Eq.~\ref{Eq:Unc_T_Ave}. We emphasized the importance of this metric to students but left them the autonomy to balance $M_{cycles}$ and $N_{trials}$, fostering independent strategic thinking.

We collected 578 individual student records spanning multiple rounds of the pendulum experiments. Since students mostly worked in pairs (very rarely solo or trios), we collapsed their data by averaging their results. Most completed at least two rounds of experiments, with the majority completing three. Only a small number proceeded to four or more rounds. For our analysis, we focused only on the first three rounds, reflecting the large majority of the data. These steps reduced the number of records to 288, each representing one pair of students per experimental round. This consolidation process focused on the type of apparatus used (simulation or physical), values of $M_{cycles}\sqrt{N_{trials}}$, standard deviations of timing measurements, and t$^\prime$-scores.

To simplify the inclusion of the standard deviation of their timing measurements $\sigma(t)$, we combined the standard deviations for the two pendulum release angles, $10^\circ$  and $20^\circ$, by taking the average of the variance for the two cases.
All statistical tests in our study were conducted with proper assumption testing and outlier handling to ensure robustness ~\cite{StatsGuide}. Univariate and multivariate outliers were identified and excluded, enhancing the reliability of our Multivariate Analysis of Variance (MANOVA) results. Further details on assumption testing and the outlier analysis process are available in Appendix A. The results of the MANOVA summarized in Table I indicate a significant difference in data quality between students using the physical apparatus, and students using the simulation.

\begin{table*}[htpb]
\centering
\label{tab:manova}
\setlength{\tabcolsep}{5pt} 
\begin{tabular}{ccccccc}
\toprule
\textbf{round} & \textbf{$n_{sim}$} & \textbf{$n_{phys}$} & \textbf{Pillai's Trace} & \textbf{F-statistic} & \textbf{adj. p-value} \\
\midrule
1 & 52 & 49 & 0.257 & $F(2, 98) = 16.934$ & .000 \\
2 & 55 & 51 & 0.081 & $F(2, 103) = 4.547$ & .051 \\
3 & 37 & 29 & 0.259 & $F(2, 63) = 10.987$ & .000 \\
Last & 54 & 48 & 0.100 & $F(3, 98) = 3.638$ & .062 \\
\bottomrule
\end{tabular}
\caption{Multivariate Analysis of Variance (MANOVA) tests were conducted for each of the first three rounds and the final round of data collection from each condition. The number of records for the simulation condition and the physical apparatus condition are $n_{sim}$ and $n_{phys}$, respectively. Tukey's HSD (honestly significant difference) tests were used for post-hoc comparisons to control for family-wise error rate across multiple pairwise comparisons. The analyses utilized two primary fitting parameters, $M_{cycles}\sqrt{N_{trials}}$ and the standard deviation of their timing measurements $\text{StdDev}(Time)$, with outlier filtration executed using the Mahalanobis Distance. In the final round, $L_{rounds}$—representing the variability in the number of iterations completed by students—was included as an additional parameter. Results indicated a statistically significant difference in data quality between simulations and physical pendulum setups. Detailed post-hoc analysis further exploring these parameters is provided in Table II, highlighting the distinct impacts of simulation and physical apparatus on experimental outcomes.
}
\end{table*}

To mitigate the risk of false positives from multiple comparisons, Tukey's HSD was used to adjust p-values for all post-hoc analyses following the MANOVA. This method controlled the family-wise error rate while providing pairwise comparisons between group means across rounds. The results of this post-hoc analysis are summarized in Table II. The adjusted p-values revealed statistically significant differences in the standard deviation of timing measurements ($\sigma(t)$) between the simulation and physical apparatus groups.

Students using the simulation consistently demonstrated more reproducible timing measurements, with a narrower distribution of timing variability compared to their counterparts using physical apparatuses. While students using the physical apparatus improved their reproducibility over successive rounds, they continued to exhibit greater variability than the simulation group. We attribute this important difference to the constraints in the simulation. With a physical apparatus, we typically observe students' measurement strategy involves trying to synchronize the release of the pendulum with the start of the stop watch. This is a particularly bad strategy when they choose to divide the tasks, with one student needing to react to the other student's release of the pendulum. This entails a reaction time that they can do little to mitigate. Even when one student does both, it involves a co-ordination of physical acts. Even worse, we have often observed that students working with a physical pendulum become quite fixated on just trying to improve this synchronization, at the expense of pursuing more fruitful strategies. This measurement approach generally yields lower precision and is also prone to worse systematic error. Students locked into this approach waste time and opportunity by focusing on this one narrow aspect of measurement uncertainty.

The quality metric ($M_{\text{cycles}} \sqrt{N_{\text{trials}}}$) did not show significant differences between groups in the earlier rounds of data collection. However, by the third round, the adjusted p-values indicated that the simulation group had significantly higher values compared to the physical apparatus group, with means of $68.6 \pm 6.7$ for the simulation and $48.6 \pm 5.2$ for the physical apparatus. Although the power of this comparison was modest (0.62), the results suggest a practical advantage for the simulation group in refining their data collection strategies. Combined with their consistently lower $\sigma(t)$, this improvement contributed to the simulation group's overall superior precision of the period measurements.

Finally, when comparing the number of measurement rounds completed by each group, no statistical or practical differences were found in the final round $L_{rounds}$ achieved by the students.

\begin{figure}[htpb]
  \centering
  \includegraphics[width=0.85\textwidth]{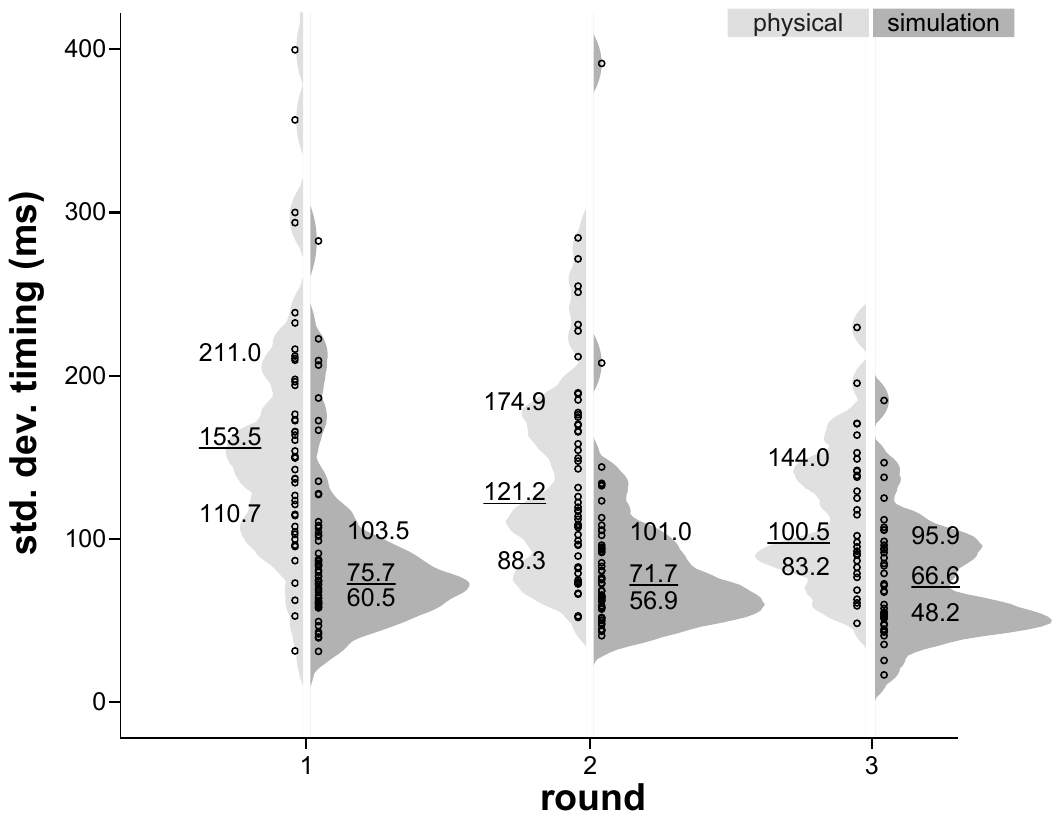}
  \caption{Bean plots depicting the evolution of the standard deviation in timing measurements as a function of round for both groups. Interpolated medians and quartile values are highlighted to underscore trends. Various statistics of these data may be seen in Table~\ref{tab:posthoc}.}
  \label{fig:ViolinPlot_StdDev}
\end{figure}

\begin{figure}[htpb]
  \centering
  \includegraphics[width=0.85\textwidth]{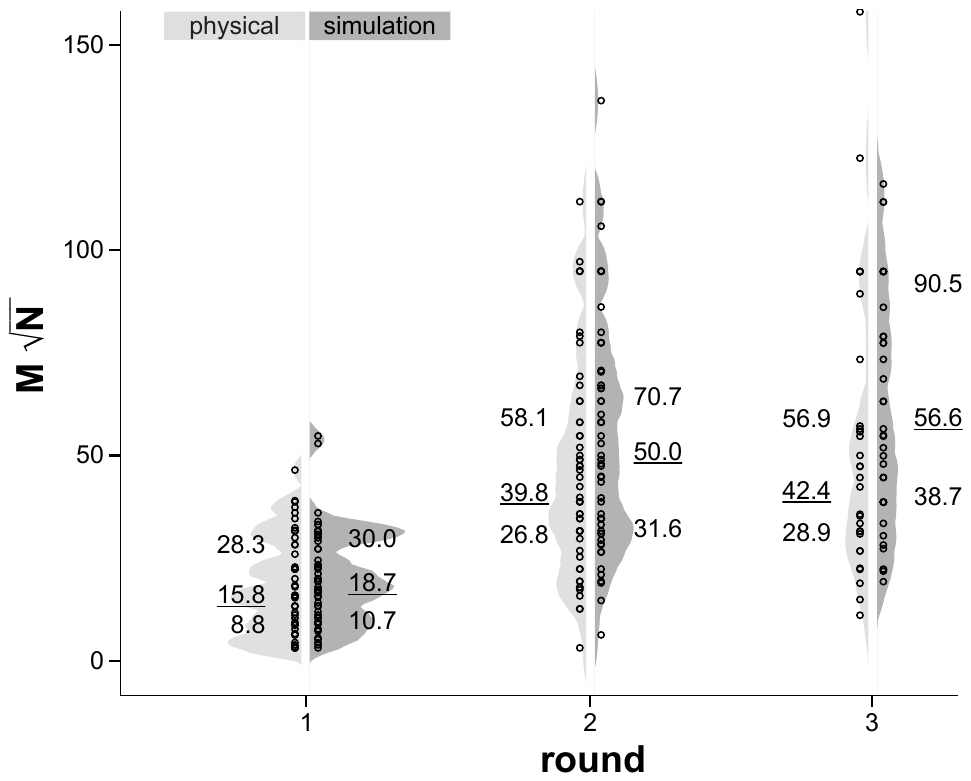}
  \caption{Bean plots depicting the evolution of the quality factor of $M_{cycles}\sqrt{N_{trials}}$ as a function of round for both groups. Interpolated medians and quartile values are highlighted to underscore trends. Various statistics of these data may be seen in Table~\ref{tab:posthoc}.}
  \label{fig:ViolinPlot_MrootN}
\end{figure}

\begin{table*}[htpb]
\centering

\setlength{\tabcolsep}{5pt} 
\begin{tabular}{cccccccc}
\toprule
\textbf{Round} & \textbf{n$_{sim}$} & \textbf{n$_{phys}$} & \textbf{mean$_{sim}$} & \textbf{mean$_{phys}$} & \textbf{difference} & \textbf{F-statistic} & \textbf{adj. p-value} \\
\midrule
\multicolumn{8}{c}{\textbf{$M\sqrt{N}$}} \\
\midrule
1 & 52 & 49 & 20.4 $\pm$ 1.6 & 18.5 $\pm$ 1.6 & 1.9 & $F(1, 99) = 0.665$ & .417 \\
2 & 55 & 51 & 55.2 $\pm$ 3.8 & 45.7 $\pm$ 3.5 & 9.4 & $F(1, 104) = 3.236$ & .075 \\
3 & 37 & 29 & 68.6 $\pm$ 6.7 & 48.6 $\pm$ 5.2 & 20.0 & $F(1, 64) = 5.129$ & .027 \\
Last & 54 & 48 & 65.5 $\pm$ 4.5 & 53.4 $\pm$ 4.7 & 12.0 & $F(1, 100) = 3.401$ & .068 \\
\midrule
\multicolumn{8}{c}{\textbf{$\sigma(t)~(ms)$}} \\
\midrule
1 & 52 & 49 & 91 $\pm$ 14 & 193 $\pm$ 15 & -102 & $F(1, 99) = 24.819$ & .000 \\
2 & 55 & 51 & 105 $\pm$ 13 & 137 $\pm$ 14 & -32 & $F(1, 104) = 3.014$ & .085 \\
3 & 37 & 29 & 74 $\pm$ 7 & 115 $\pm$ 7 & -41 & $F(1, 64) = 16.872$ & .000 \\
Last & 54 & 48 & 88 $\pm$ 8 & 121 $\pm$ 9 & -33 & $F(1, 100) = 7.730$ & .060 \\
\midrule
\multicolumn{8}{c}{\textbf{Round}} \\
\midrule
Last & 54 & 48 & 2.70 $\pm$ 0.07 & 2.58 $\pm$ 0.08 & 0.12 & $F(1, 100) = 1.273$ & .262 \\
\bottomrule
\end{tabular}
\caption{In the post-hoc analyses following the MANOVA, Tukey's HSD tests were used to further investigate differences between the simulation and physical apparatus conditions. Tukey's HSD adjusts for multiple comparisons by controlling the family-wise error rate, providing pairwise comparisons between group means while mitigating the risk of Type I errors. Statistically significant differences were observed in the standard deviation of timing measurements $\sigma(t)$ across all rounds, indicating a narrower distribution of data for the simulation group, suggesting more consistent measurement techniques. While no difference for the quality factor $M_{cycles}\sqrt{N_{trials}}$ appeared in the initial rounds, by round three a significant difference emerged, suggesting that simulation users adopted more effective data collection strategies.
}
\label{tab:posthoc}
\end{table*}

\clearpage

\subsection{\label{subsec:disagreement} Perceptions of Disagreement}

The implications of obtaining higher precision, and consequently, a higher value of the t$^\prime$-score, are important to the lab goals, with t$^\prime$-score greater than 3 being the threshold for a significant difference between the two pendulum periods. For the data presented in Fig. \ref{fig:tscore_evolution}, 52\% of the student groups using the physical apparatus arrive at a t$^\prime$-score that is 3 or greater, whereas 78\% of those using a simulation reach that level.

Beyond these data, we explore how well students using either physical or simulated pendulums recognized discrepancies between their measured periods of pendulum cycles, by analyzing responses from an anonymous, post-lab survey. Students were asked for their measured values of the period, with uncertainty, measured at $10^\circ$ and $20^\circ$. Of the individual students responding, 66 used a physical pendulum and 75 used a simulation. As shown in Table \ref{table:disagreement}, 53 \% of the students using physical apparatus reported a t$^\prime$-score of 3 or greater, but 75\% got to this level using the simulation. This improved outcome is in close accord with the student data presented in Fig. \ref{fig:tscore_evolution}.

When asked about their perception of a disagreement between periods, two interesting results emerge. Only 39\% of students using the physical apparatus indicated they perceived slight or strong disagreement, compared to 55\% in the simulation group, suggesting that the higher t$^\prime$-scores of the latter group are beneficial. However, focusing on students whose t$^\prime$-scores exceeded 3, indicating they had data to support a conclusion of disagreement between measurements, 74\% using physical apparatus and 73\% using the simulation acknowledged disagreement. This similarity in responses suggests that, when the data clearly indicated a disagreement, students from both conditions were equally likely to interpret their data correctly, regardless of the experimental setup, but that roughly a quarter of the students are reluctant to conclude disagreement, even when their data support that conclusion.

\begin{table}[htpb]

\label{tab:disagreement_perception}
\begin{tabular}{cccccccc}
\hline
\textbf{Condition} & \begin{tabular}[c]{@{}c@{}} \textbf{Number} \\ \textbf{reporting} \\ \textbf{t$^\prime$} \\ \end{tabular} & \begin{tabular}[c]{@{}c@{}} \textbf{Number} \\ \textbf{reporting} \\ \textbf{t$^\prime \geq$ 3} \\ \end{tabular} & \begin{tabular}[c]{@{}c@{}} \textbf{Number} \\ \textbf{reporting} \\  \textbf{t$^\prime \geq$ 3} and \\ \textbf{disagreement} \\ \end{tabular} \\ \hline
Physical       & 66                   & 35   &   26               \\ \hline
Simulation     & 75                   & 56   & 41            \\ \hline
\end{tabular}
\caption{Perception of Disagreement by Condition and t$^\prime$-score. 75\% of the students using the simulation reach \textbf{t$^\prime \geq$ 3}, but only 53\% of students using the physical apparatus reach that level. As a consequence, 55 \% of the students using the simulation report a disagreement between the periods, but only 39 \% of students using the physical apparatus come to this conclusion.}
\label{table:disagreement}
\end{table}

\subsection{\label{SubSec:perspectives} Student Perspectives}

The anonymous post-lab survey offers further insight into the students' perception of the lab.

\subsubsection{\label{SubSubSec:Trust} Trust in their equipment}
Survey question 4 was an open-ended question about the student's associated trust with the simulated pendulum versus the physical one. Coding for this open-ended question consisted of two authors (JB and JD) undergoing several of several rounds of coding and discussions, eventually deciding on three categories with 100\% agreement between the coders. These three categories were:

\begin{enumerate}
\item Trust simulated pendulums over physical pendulums
\item Trust physical pendulums over simulated pendulums
\item Other
\end{enumerate}
 The results of question 4 (Table~\ref{Table:trust}) showed that both groups overwhelmingly claim to trust the results from simulation labs more than their analogous physical labs, with no difference in levels of trust between the two groups.

\begin{table}[htpb]
\begin{tabular}{ccccc}
\hline \hline
Type & Count &
  \begin{tabular}[c]{@{}c@{}c@{}}Trust \\ Sim $\%$ \end{tabular} &
  \begin{tabular}[c]{@{}c@{}c@{}}Trust \\ Phys $\%$ \end{tabular} &
  \begin{tabular}[c]{@{}c@{}c@{}}Trust \\ Other $\%$ \end{tabular} \\
  \hline \\
Phys	&$	66	$&$	68.2	$&$	13.6	$&$	18.2	$ \\
Sim	    &$	76	$&$	71.1	$&$	7.9	    $&$	21.1	$ \\
\hline \hline
\end{tabular}
\caption{Distributions from the open-ended survey question 4 which asks students which type of set-up they would trust more.  In either instance, more students trust a simulation.}
\label{Table:trust}
\end{table}

 \subsubsection{\label{SubSubSec:Experience} Student Experience}
As for the remainder of the survey, two questions specifically address the student experience with their equipment/simulation. Question 6 asked how difficult they found it relative to previous labs. Students using the simulation tilted more towards saying they found it less difficult or non-intuitive than previous labs. Related to this, question 9 asked how their set-up affected lab practices such as performing improvements/iterations, doing uncertainties, and performing analysis. In response, the students using the physical apparatus said they could do about the same amount, but students using the simulation indicated they could do a bit more than usual.

For questions 5 and 7, there was little difference between students using the simulations versus those using the physical apparatus. In response to question 5, both groups leaned towards saying their experience in the labs was somewhat better than previous labs. In response to question 7, both groups stated that they discussed things with other students about as often in previous labs, suggesting the simulation did not interfere with other desired lab behaviours.

Question 8 asked students if they would recommended using simulations in future iterations of the course. Students who used simulations reported wanting \textquotedblleft slightly more simulations than physical labs in the future," while those who used physical pendulums reported wanting an \textquotedblleft even mix (of) simulations and regular labs in the future".

\begin{table}[ht]
\centering
\caption{Comparison of Interpolated Medians and (Dispersion Indices) for Physical and Simulation Groups. The survey question wording can be found in Appendix B.}
\label{tab:Comparison}
\begin{tabular}{@{}cccc@{}}
\toprule
\textbf{Question} & \textbf{Physical} & \textbf{Simulation} \\ \midrule
Q5                & 3.88 (0.53) & 3.77 (0.50) \\
Q6                & 3.43 (0.47) & 4.21 (0.50) \\
Q7                & 3.14 (0.52) & 2.95 (0.54) \\
Q8                & 3.12 (0.56) & 3.76 (0.59) \\
Q9                & 3.07 (0.45) & 3.88 (0.54) \\
\bottomrule
\end{tabular}
\end{table}

\section{\label{Sec:Conclusion} Conclusions}
The study presented here focuses on a pivotal experiment in a first-year inquiry lab. The overall lab course is designed to build student abilities in experimentation and the use of statistical methods, and to form strong habits of mind in their approach to quantitative experimental questions. The importance of the particular experiment presented here relies on the impact of students experiencing conflicting results and the failure of a model as they acquire increasingly precise data. In this regard, it is critical that as many students as possible are able to navigate this path to its end point, while giving them the time and free agency to get there on their own. In service of that goal, we have compared the use of a physical apparatus and a simulation. While both are carefully designed, we have shown that the simulation provides affordances and constraints that lead to superior student outcomes.

The simulation was carefully tuned so that amplitude decay matched that of the physical pendulum, meaning the two conditions were equivalent in the physical quantity most relevant to measurement quality. This makes the logistics of the timing method the most compelling explanation for the difference in outcomes we observe. In particular, the design of the simulation prevents students from getting trapped in a timing technique that is most susceptible to systematic and random errors associated with human reaction time. Students working in pairs with a physical apparatus often start on a measurement method where one operates the timer and one releases the pendulum. This is inevitably affected by a simple reaction time problem. Even worse, rather than try something fundamentally different in subsequent attempts to improve their measurements, students can get stuck in fruitlessly trying to improve their synchronization. Even when a single student does both, it is a complex act of coordination. In the simulation, they are forced to set the pendulum swinging, then just choose an easily-anticipated point in the swing to start and stop the timer. The students are still doing the timing measurements themselves and still have ample free agency in other experimental design choices such as where in the cycle to time from, how many cycles to count, and how many trials to average. But, they are by design prevented from getting trapped in a low-precision dead end. The result is that students using the simulation are better able to progress through rounds of iterative improvements to their measurements, and uncover the failure of the simple harmonic model of the pendulum, the desired endpoint of the experiment. With better data, a greater portion of students using the simulation arrive at the challenging conclusion we want them to reach, which is that the period of a pendulum does depend on the angle at which it is swinging.

There is often a conflict between the need to efficiently make use of class instruction and lab time, and the need to give students the opportunity to learn for themselves. This conflict runs through many learning goals and learning environments, but is especially acute in a situation where we want students to learn through authentic experimental inquiry. It is surprisingly difficult to design a physical experiment that meets these goals in limited time. A carefully designed simulation allows one to improve a student's odds of success, while still giving them the freedom to construct new knowledge for themselves. Recent comparative work on pendulum lab design across institutions frames such choices as an ``agentive trade-off'': constraining some student decisions in order to preserve or expand agency in more consequential ones~\cite{Descamps2026}. Our results provide empirical grounding for one such trade-off, where a single timer constraint removed a low-value decision from students' hands while leaving them full autonomy over all higher-order experimental choices. The ability to program in various affordances and constraints in a simulation allows one to softly guide students in productive directions.

\begin{acknowledgments}

This work has been supported by the Department of Physics \& Astronomy and by Teaching as Research (TAR) funds thanks to the Centre for the Integration of Research, Teaching, and Learning (CIRTL), both at the University of British Columbia.

\end{acknowledgments}

\bibliography{pendulum}

\appendix

\section{Student Survey}
\label{Appendix:Supplementary Materials}

\begin{figure*}[htpb]
  \centering
  \includegraphics[width=0.95\textwidth]{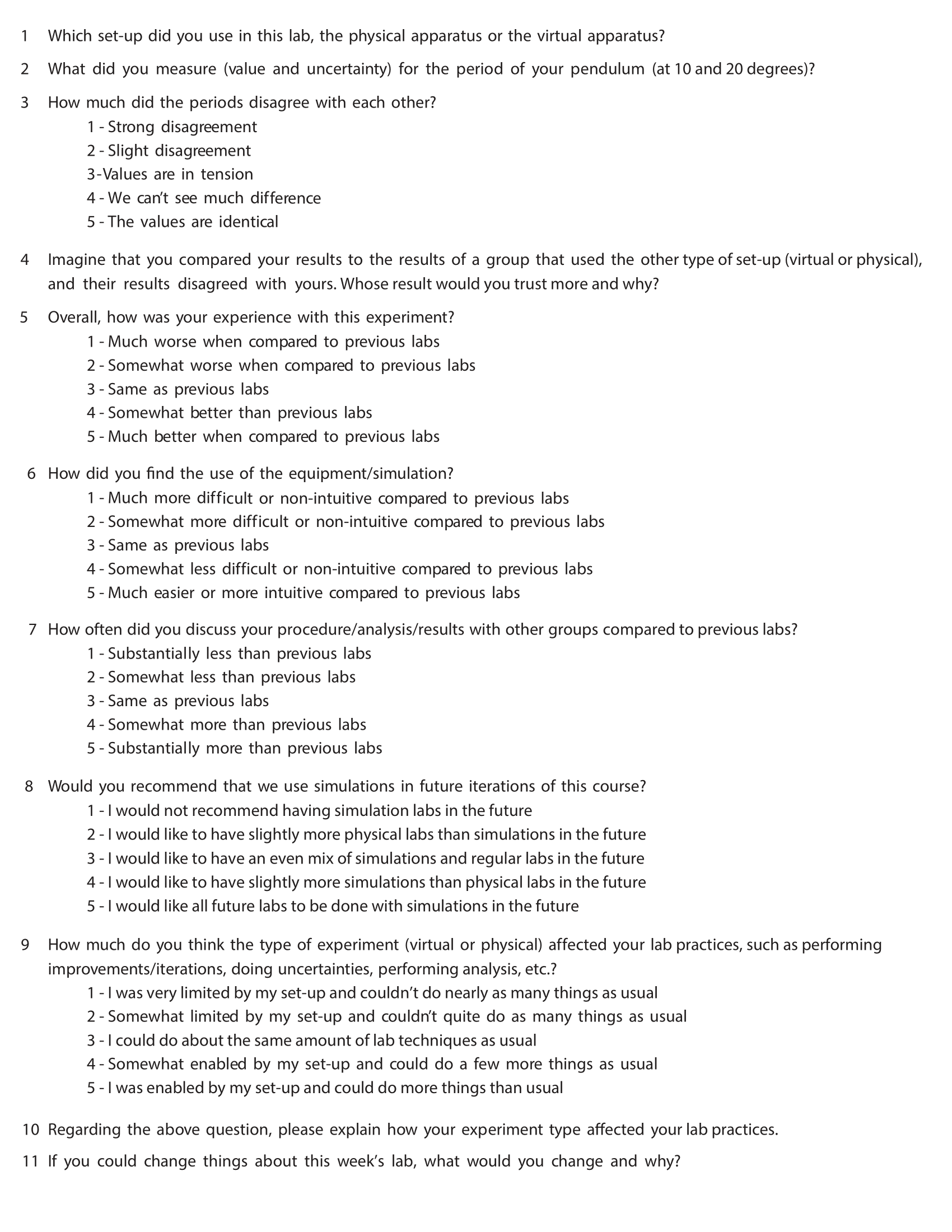}
  \caption{The post-lab anonymous survey. ($N_{Physical}=71$, $N_{Simulation}=79$, $N_{Total}=150$)}
  \label{fig:survey_questions}
\end{figure*}

The post-lab survey, presented in Fig.~\ref{fig:survey_questions}, was administered immediately after the lab session. The survey included Likert-type questions and open-text responses.

\subsection{Coding Open Responses}
For open-text question 4, two authors independently coded responses, reconciling any disagreements through discussion. The coding process involved initial categorization, refining categories, and achieving consensus on the final coding.

\clearpage
\section{Analyses Background}
\label{Appendix:Analyses_Background}

This appendix provides additional detail on the statistical methods~\cite{StatsGuide} and assumptions underlying our analysis. We include it here to support transparency and reproducibility for readers who may wish to evaluate or replicate our approach.

\textbf{Outlier Detection:}
\begin{itemize}
  \item \textit{Univariate outliers} were identified using boxplots.
  \item \textit{Multivariate outliers} were flagged using the Mahalanobis distance, with a threshold of $p = 0.001$.
  \item All identified outliers were excluded from subsequent analyses to improve the robustness of the MANOVA results.
\end{itemize}

\textbf{Assumption Testing:}
We verified standard assumptions for MANOVA, including normality, independence, and homogeneity of variances. The details of these checks follow the procedures recommended in established statistical guides.

\textbf{Statistical Tests Used:}
Multivariate Analysis of Variance (MANOVA) was used to test whether students in the simulation and physical apparatus groups differed across multiple continuous dependent variables, such as measurement variability and quality metrics. This method allowed us to assess interactions across multiple related outcomes in each experimental round.

\textbf{Post-hoc Comparisons:}
To control for family-wise error when making multiple pairwise comparisons, we used Tukey's Honestly Significant Difference (HSD) test. Compared to the Bonferroni correction, which can be overly conservative and reduce statistical power, Tukey’s HSD provides a more balanced approach when testing multiple means. This choice was guided by its appropriateness for the number of comparisons and group sizes in our study.

\end{document}